\documentclass[fleqn,usenatbib]{mnras}

\usepackage{newtxtext,newtxmath}
\usepackage[T1]{fontenc}

\DeclareRobustCommand{\VAN}[3]{#2}
\let\VANthebibliography\thebibliography
\def\thebibliography{\DeclareRobustCommand{\VAN}[3]{##3}\VANthebibliography}

\usepackage{graphicx} % Including figure files
\usepackage{multirow}
\renewcommand\ion[2]{#1\,\textsc{\lowercase{#2}}}  % ionization states
\title[The remnant of V5668 Sgr]{Optical and NIR data and modelling of nova V5668 Sgr}

\author[L. Takeda et al.]{L. Takeda,$^{1}$\thanks{E-mail: larissa.takeda@usp.br}
M. Diaz,$^{1}$
R. D. Campbell,$^{2}$
J. E. Lyke,$^{2}$
S. S. Lawrence,$^{3}$
J. D. Linford,$^{4}$
\newauthor{K. V. Sokolovsky$^{5,6}$}\\
$^{1}$Universidade de S\~ao Paulo, Instituto de Astronomia, Geof\'isica e Ci\^encias Atmosf\'ericas,  Rua do Mat\~ao, 1226 , Sao Paulo, SP 05508-900, Brazil\\
$^{2}$W. M. Keck Observatory, 65-1120 Mamalahoa Hwy., Kamuela, HI 96743, USA\\
$^{3}$Dept. of Physics \& Astronomy, 151 Hofstra University, Hempstead, NY 11549, USA\\
$^{4}$National Radio Astronomy Observatory, Socorro, NM 87801, USA \\
$^{5}$Center for Data Intensive and Time Domain Astronomy, Department of Physics and Astronomy, Michigan State University, 567 Wilson Rd, East Lansing,\\
MI 48824, USA\\
$^{6}$Sternberg Astronomical Institute, Moscow State University, Universitetskii~pr.~13, 119992~Moscow, Russia}

\date{Accepted 2022 January 07. Received 2021 December 07; in original form 2021 October 09}

\pubyear{2021}

\begin{document}
\label{firstpage}
\pagerange{\pageref{firstpage}--\pageref{lastpage}}
\maketitle

\begin{abstract}
We present HST optical images, Keck-OSIRIS NIR IFS data cubes and Keck-NIRC2 NIR images of nova V5668 Sgr from 2016 to 2019. The observations indicate enhanced emission at the polar caps and equatorial torus for low ionization lines, and enhanced high ionization emission lines only at the polar caps. The radial velocities are compatible with a homogeneous expansion velocity of v=590~km\,s$^{-1}$ and a system inclination angle of 24$\degr$. These values were used to estimate an expansion parallax distance of 1200 $\pm$ 400 pc. The NIRC2 data indicate the presence of dust in 2016 and 2017, but no dust emission could be detected in 2019. The observational data were used for assembling 3D photoionization models of the ejecta. The model results indicate that the central source has a temperature of $1.88\times10^{5}$~K and a luminosity of $1.6\times10^{35}$ erg\,s$^{-1}$ in August of 2017 (2.4 years post eruption), and that the shell has a mass of $6.3\times10^{-5}$~M$_{\sun}$. The models also suggest an anisotropy of the ionizing flux, possibly by the contribution from a luminous accretion disc. 

\end{abstract}


%%%%%%%%%%%%%%%%%%%%%%%%%%%%%%%%%%%%%%%%%%%%%%%%%%

%%%%%%%%%%%%%%%%% BODY OF PAPER %%%%%%%%%%%%%%%%%%

\section{Introduction}

Nova eruptions are powered by nuclear fusion that ignites at the bottom of a hydrogen-rich shell on the surface of an accreting white dwarf in a binary system \citep{2008clno.book.....B,1972ApJ...176..169S,2016PASP..128e1001S}. The eruption leads to ejection of this shell mixed with the white dwarf material at velocities of hundreds to thousands of km\,s$^{-1}$. The ejected gas often presents complex structures due to the combination of the velocity distribution, instabilities and nonuniform mass loss. Clumps, asymmetries and bipolar structures are commonly observed in nova shells \citep{RibeiroRsOph,2009AJ....138.1541M} and may play a key role in the evaluation of the physical and chemical parameters of novae. Although condensed regions of gas in the shell are necessary to explain the range of ionization states observed in nova spectra, most modelling of nova shells adopt simplistic density profiles, such as power-law functions. The discrepancy of the modelled and observed mass distributions can lead to inaccuracies in derived shell masses and chemical abundances, among other properties. Accurate estimates of the ejecta masses and abundances are needed to understand how novae contribute to chemical evolution of the Galaxy, especially for such species as $^{7}$Li \citep{2015ApJ...808L..14I} and $^{26}$Al \citep{1997ApJ...479L..55J}, and if white dwarfs in nova systems can evolve to type Ia supernovae \citep{Nomoto}. 

Considering the expansion velocities of hundreds to thousands of km\,s$^{-1}$, the ejecta of many novae can be resolved with high-resolution imaging a few years after eruption. The study of resolved nova shells can provide valuable insights into the process of nova eruption and the development of the observed structures in the ionized shell. The formation and evolution of clumps and their relation to shocks along with the overall geometry and mass distribution of the ejecta are just a few of the outstanding questions in nova eruption physics \citep{2020MNRAS.491.4232S,2020ApJ...905...62A}. Likewise, the different morphologies frequently observed for distinct lines in spectral data cubes require investigations to clarify if their origin lies mostly in a density gradient (derived from anisotropic ejecta distribution, shocks and other processes), an abundance gradient, or an anisotropy of the ionizing photon field.

The data cubes produced by Integral Field Spectroscopy (IFS) are powerful observational tools to gather multiple types of information from different sections of the shell simultaneously \cite[for example see][]{2009AJ....138.1090L}. The use of spatially resolved spectroscopy allied to photoionization models has proved to provide deeper insights into the nova ejecta structure compared to the unidimensional line-of-sight treatment \citep{1992MNRAS.258P...7E,TakedaV723Cas}.

V5668 Sgr (Nova Sagittarii 2015b) was detected on March~15, 2015 by \citep{CBET} and was extensively observed across the electromagnetic spectrum. In the early phase of the eruption, the nova spectra presented $^{7}$Be lines \citep{2016ApJ...818..191T}, revisiting the discussion of novae as possible important contributors to Galactic lithium. V5668 Sgr was one of the first few novae detected in high-energy ($> 100$~MeV) gamma-rays, in the early phase of eruption \citep{2016ApJ...826..142C}. The link between gamma-ray and optical variabilities observed in some novae and the presence of gamma-ray emission accompanying the optical variability in V5668 Sgr are believed to result from internal shocks in colliding shells of multiple ejecta \citep{2017NatAs...1..697L}. Dust formation was detected around $80$ days after the eruption, reaching maximum grain condensation around day $100$, followed by grain (at least partial) destruction by day $200$. Concomitant with the dust observation, during the period of $170-240$ days, soft ($0.3-1$~keV) and hard ($2-10$~keV) X-ray emission was also detected \citep{2018ApJ...858...78G}. In later stages of the eruption, ALMA observations showed that the gas was condensed in small ($\leq 10^{15}$~cm) clumps in the shell \citep{ALMA}. All these events make V5668 Sgr an especially interesting target, that can be used to study possible correlations between different time-lapsed processes.

\begin{figure*}
    \centering
    \includegraphics{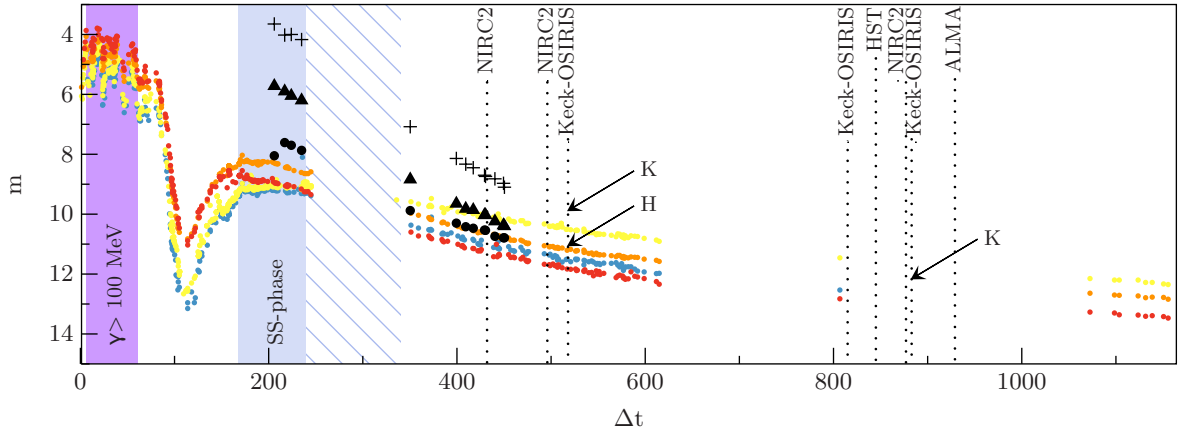}
    \caption{Light curve of V5668 Sgr eruption with t$_0 = 2015$ March 15. The blue, yellow, orange and red dots correspond to \textit{B, V, R} and \textit{I} magnitudes from AAVSO archive, with errors not greater than 0.1. The purple region corresponds to the period with gamma-ray detection \citep{2016ApJ...826..142C}. The blue region corresponds to the period of super-soft phase \citep{2018ApJ...858...78G}, that ended within the dashed blue region. The \textit{K} and \textit{H} indicated by arrows correspond to 2MASS \textit{Ks} and \textit{H} magnitudes derived from our flux calibrated Keck-OSIRIS NIR spectra, with errors of 0.3 (see section \ref{sec:keck}). The black dots, triangles and crosses correspond to \textit{J}, \textit{H} and \textit{K} magnitudes with maximum error of 0.2 \citep{2018ApJ...858...78G}. Our HST, Keck-NIRC2 and Keck-OSIRIS observations are marked on the light-curve, as well as ALMA observations \citep{ALMA}.}
    \label{fig:lc}
\end{figure*}

In this paper, we present \textit{Hubble Space Telescope} (HST) optical images and near-infrared Keck-OSIRIS IFS data and Keck-NIRC2 imaging of V5668 Sgr, observed in the period of $431-882$ days after detection. Along with the analysis of the imaging and spectroscopic evolution, we provide estimates of the physical and chemical properties obtained through 3D photoionization models. The optical and NIR light curves of V5668 Sgr are displayed in Figure \ref{fig:lc}, with indications of our observations and the previous cited events.

\section{HST data}

\begin{figure}
    \centering
    \includegraphics{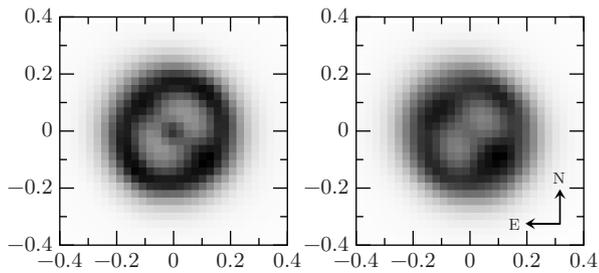}
    \caption{HST images in filters F657N (left) and F502N (right). The angular radii are given in arcsec relative to the central source.}
    \label{fig:hst}
\end{figure}

We obtained \textit{Hubble Space Telescope} Wide Field Camera 3 images in the H$\alpha+[$\ion{N}{II}] (F657N) and [\ion{O}{iii}] 5007 \AA\ (F502N) narrow filters, displayed in Figure \ref{fig:hst}. The observation was made at $\Delta$t = 844 days, on 2017 July 06 as part of HST-GO-14787. The total fluxes for each filter, including the continuum emission and central source emission, are f$_{\rm F657N}=2.6\times10^{-11}$ and f$_{\rm F502N}=1.1\times10^{-10}$~erg\,cm$^{-2}$\,s$^{-1}$. These fluxes were measured in the pipeline-processed drizzled images and are not dereddened. Direct measurements of field stars in the images at native resolution indicate that the stellar PSFs have FWHMs of 79 and 83 milliarcseconds in the F657N and F502N filters, respectively.

In both filters the remnant appears as a boxy ellipsoid with a major axis aligned SE-NW and a minor axis running NE-SW that appears to terminate in two knots of brighter emission. We interpreted these knots as polar emission, suggesting distinct densities in the equatorial and polar regions or an anisotropy of the ionizing field, or both. Additional NIR IFS analysis provides further support for the hypothesis of anisotropic ionization, which will be discussed in section \ref{sec:discussion}. 

By fitting gaussian functions to the peaks of the radial profiles extracted along the minor and major axes, we find the remnant diameters to be of 0.28 x 0.34 arcsec in F657N and 0.25 x 0.32 arcsec in F502N. The full width at 10 per cent maximum (FW0.1M) extends fairly circularly to 0.6 arcsec in all directions. The central point source is distinctly stronger in F657N, since forbidden transitions do not occur in the dense central source and the continuum emission is weak on the observing dates.

\section{Keck-OSIRIS data} \label{sec:keck}

Spatially resolved spectra of V5668 Sgr were obtained with Keck-OSIRIS IFU spectrograph on three different epochs. For all dates, we used a laser guide star adaptive optics (LGSAO) system to achieve the image quality of $\sim50$~mas, sampled at 35 mas per IFU lenslet. On 2016 August 13, we obtained data in \textit{H} and \textit{K} bands, with 2 exposures of 60~s for \textit{H} broad band, and 2 exposures of 60~s for each \textit{K} moderate band filter (Kn1, Kn2, Kn3, Kn4 and Kn5). On 2017 June 06, we observed the nova using Kn2 and Kn3, with 2 exposures of 300~s for each filter, and on 2017 August 13, we used Kn1, Kn2, Kn3 and Kn5 filters, with 2 exposures of 300~s for each filter. The spectral resolution is R $\sim$3800 or $\sim$79 km\,s$^{-1}$ in the \textit{K} band. The data were reduced with the OSIRIS Data Reduction Pipelines.

\begin{figure*}
    \centering
    \includegraphics{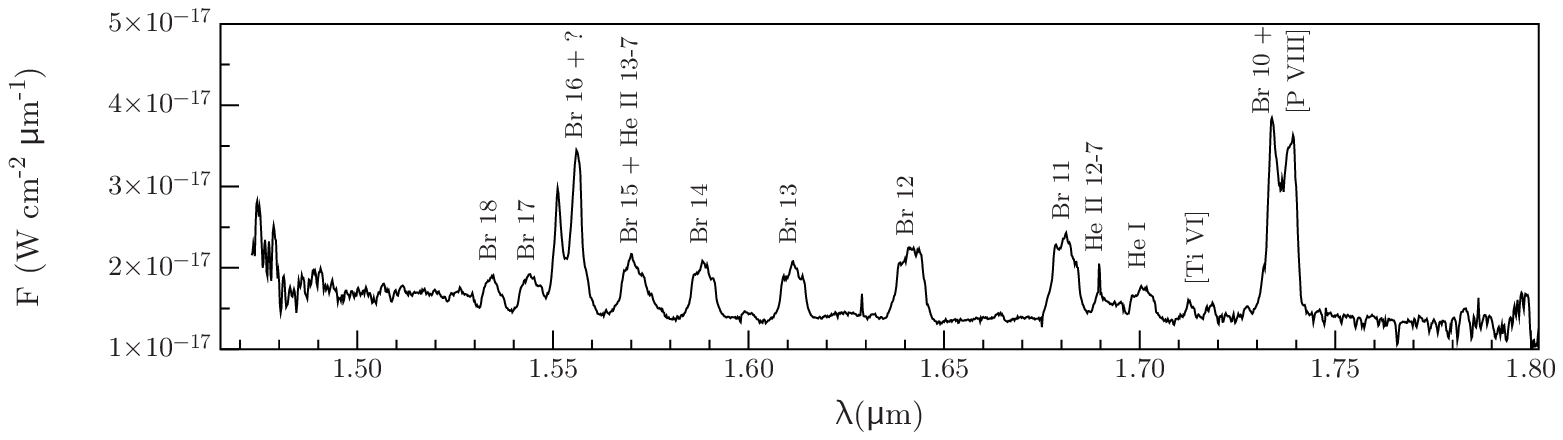}
    \caption{V5668 Sgr integrated spectrum in \textit{H} band obtained on August 13, 2016.}
    \label{fig:h_spec}
\end{figure*}

\begin{figure*}
    \centering
    \includegraphics{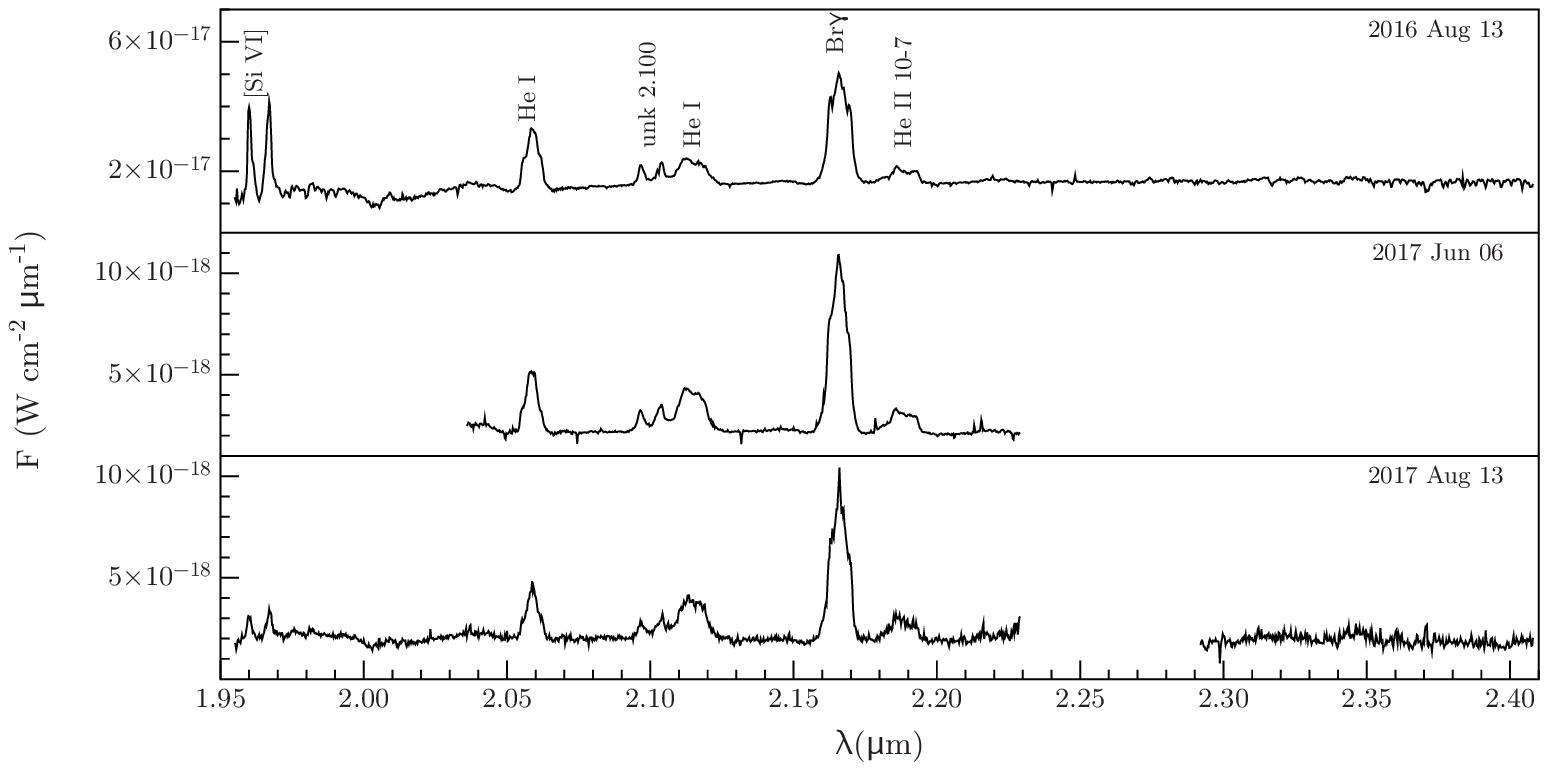}
    \caption{Spectral evolution of V5668 Sgr shell integrated spectra in \textit{K} band.}
    \label{fig:k_spec}
\end{figure*}

The telluric correction was performed using HD 189920 standard in the first two epochs, and using HIP 106329 in the third. The same standards were used to calibrate the flux of Keck-OSIRIS spectra based on their \textit{H} and \textit{Ks} 2MASS magnitudes \citep{2003yCat.2246....0C}. Considering the sensitivity differences between the narrow band filters and the errors in the telluric magnitude, we expect the errors of our derived magnitudes for V5668 Sgr to be in the order of 0.3 mag. The shell integrated flux calibrated spectra of V5668 Sgr are displayed in Figures \ref{fig:h_spec} and \ref{fig:k_spec}, for \textit{H} and \textit{K} bands respectively. The integrated \textit{H} and \textit{K} magnitudes of V5668 Sgr are displayed in Figure \ref{fig:lc} where they can be compared to previous \textit{JHK} observed magnitudes. For further analysis, we will apply a reddening correction assuming \textit{E(B-V)}=0.2, on the basis of \textit{E(B-V)}=0.2-0.3 suggested in the literature \citep{2011ApJ...737..103S,2015ATel.8275....1K,2018ApJ...858...78G}.

The integrated spectra show the Brackett series along the \textit{H} and \textit{K} bands, as well as \ion{He}{i} and \ion{He}{ii} transitions and a few forbidden lines. As has been observed in previous V5668 Sgr integrated spectra \citep{2018ApJ...858...78G,2017ATel10557....1W}, there are two types of line profiles: the the combination of a central peaked emission with a double peaked feature, and the purely double peaked lines. The first line structure is seen in the Brackett and \ion{He}{i} lines that are not blended with forbidden lines. The second one is attributed to the forbidden and higher-ionization transitions, such as [\ion{Si}{vi}], [\ion{Ti}{vi}], [\ion{P}{viii}] and the unidentified lines at 2.100~\micron\ and 1.555~\micron, and possibly to the \ion{He}{ii} line at 2.189~\micron, although the asymmetry of this feature and its high flux relative to the recombination flux expected from \ion{He}{i} 2.058~\micron\ intensity suggest the presence of another transition with close wavelength. It is interesting to notice the absence of the [\ion{Ca}{viii}] emission line, which was present in earlier ($\Delta$t $\sim350$~days) spectra \citep{2018ApJ...858...78G}. The [\ion{Ca}{viii}] line requires a slightly lower ionization energy (147.3~eV) than [\ion{Si}{vi}] (205.3~eV), that is still observed in the spectra, but a lower transition probability (A=0.72~s$^{-1}$) than [\ion{Si}{vi}] (A=2.38~s$^{-1}$) \citep{NIST_ASD}. The narrow interval of possible physical parameters of the ionizing source in order to reproduce the observed spectrum turns out to be an important constraint in the photoionization modelling of the shell.

\begin{figure}
    \centering
    \includegraphics[width=8cm]{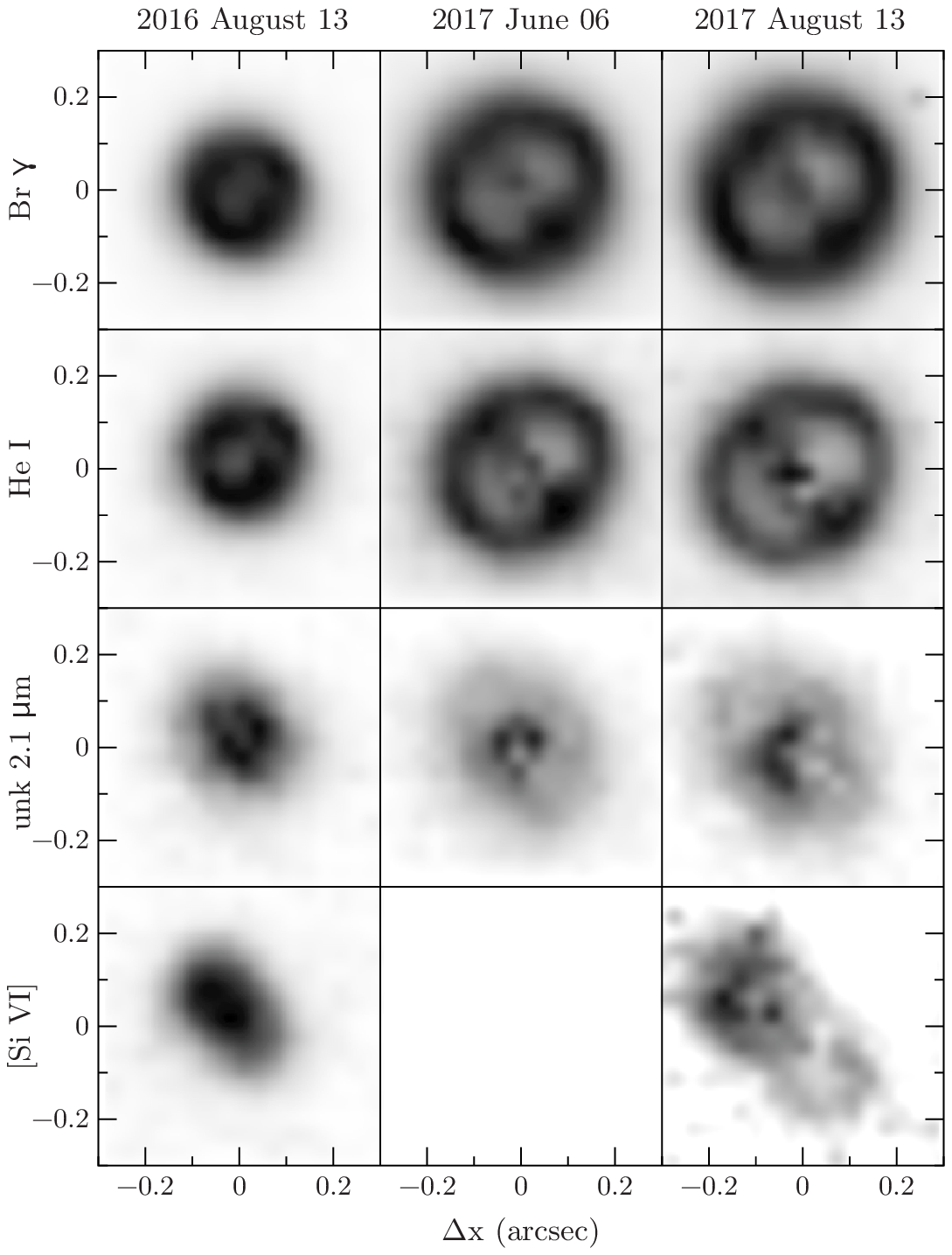}
    \caption{V5668 Sgr shell evolution for Br$\gamma$, \ion{He}{i} $2.058$~\micron, unidentified line at $2.100$~\micron\ (noted as `unk') and [\ion{Si}{vi}] lines. Both axes represent the position related to the center of the shell measured in arcseconds. The continuum was subtracted in all images. The images are shown with north-up and east-left.}
    \label{fig:osirisim}
\end{figure}

\begin{figure*}
    \centering
    \includegraphics[width=12cm]{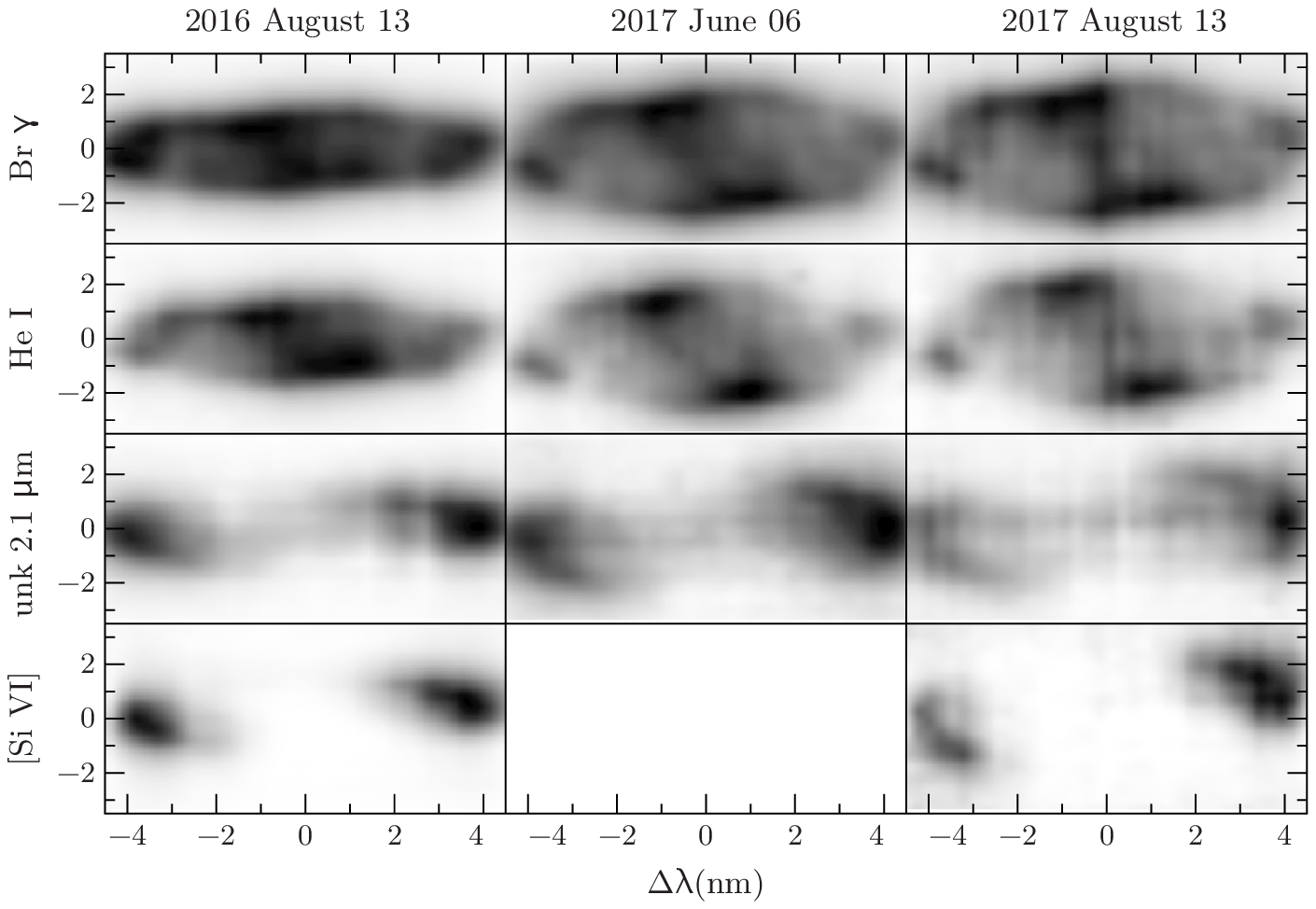}
    \caption{Temporal evolution of V5668 Sgr shell projection along RA axis for Br$\gamma$, \ion{He}{i} $2.058$~\micron, unidentified line at $2.100$~\micron\ (noted as `unk') and [\ion{Si}{vi}] lines. The y-axis represents the position (in Declination axis) related to the center of the shell measured in arcseconds and the x-axis is the wavelength related to the line center. The continuum was subtracted in all images, but it is possible to notice a residual track of the continuum in the blended unidentified line at 2.1~\micron\ emission.}
    \label{fig:osirispv}
\end{figure*}

The spatially resolved spectra allows us to identify the structures responsible for these different emission line features observed in 1D spectra. The large scale shell structure observed in the permitted lines is composed of polar caps and an equatorial torus, while the forbidden lines present only the polar caps emission. It is important to stress that although we are considering this large scale smooth geometry, ALMA high spatial resolution radio observations have shown that these structures are actually a result of an unresolved clumpy gas distribution \citep{ALMA}. The temporal evolution of the integrated images in four different emission lines are shown in Figure \ref{fig:osirisim}, with their corresponding spectral profiles in Figure \ref{fig:osirispv}. We note the similarity of the structures for the permitted transitions of Br$\gamma$ and \ion{He}{i} 2.058~\micron, that also matches the general structure of HST H$\alpha$ image. For all epochs, the \ion{He}{i} shell radii correspond to $\sim 90$ per cent of the Br$\gamma$ radii, which can be explained by the difference in ionization energies. 

For the unidentified line at 2.100~\micron\ and the [\ion{Si}{vi}] line, the polar emission becomes evident through the spectral profile. The absence of an equatorial torus and the stronger polar emission when compared to the other lines may imply the presence of an anisotropic ionizing source that enhances the ionization at polar caps. In this scenario, the [\ion{O}{iii}] HST image presents strong polar emission and an equatorial torus because [\ion{O}{iii}] has an intermediate ionization (54.9~eV) between \ion{He}{i} and [\ion{Si}{vi}]. For [\ion{Si}{vi}], with much higher ionization energy (205.3~eV), the polar caps emission completely dominate. Since the unidentified line at 2.100~\micron\ has an emission profile similar to the [\ion{Si}{vi}], it probably comes from a highly ionized ion, with energy significantly higher than 55~eV. A similar feature was noticed in the shell of V723 Cas \citep{2009AJ....138.1090L}, in which the [\ion{Al}{ix}]\ emission presented a bipolar shape, while lower ionization transitions, such as [\ion{Si}{vi}]\ and [\ion{Ca}{viii}], presented both equatorial torus and polar caps structures. In the case of V723 Cas, the Br$\gamma$ emission was too faint and dominated by the central source contribution, thus the shell emission morphology could not be defined. Regarding the shell size, the mean angular radius of Br$\gamma$ brighter structures is $180$~mas. The HWHM is $260$~mas and the total radius is $350$~mas.

\section{NIRC2 data}
\begin{figure*}
    \centering
    \includegraphics[width=12cm]{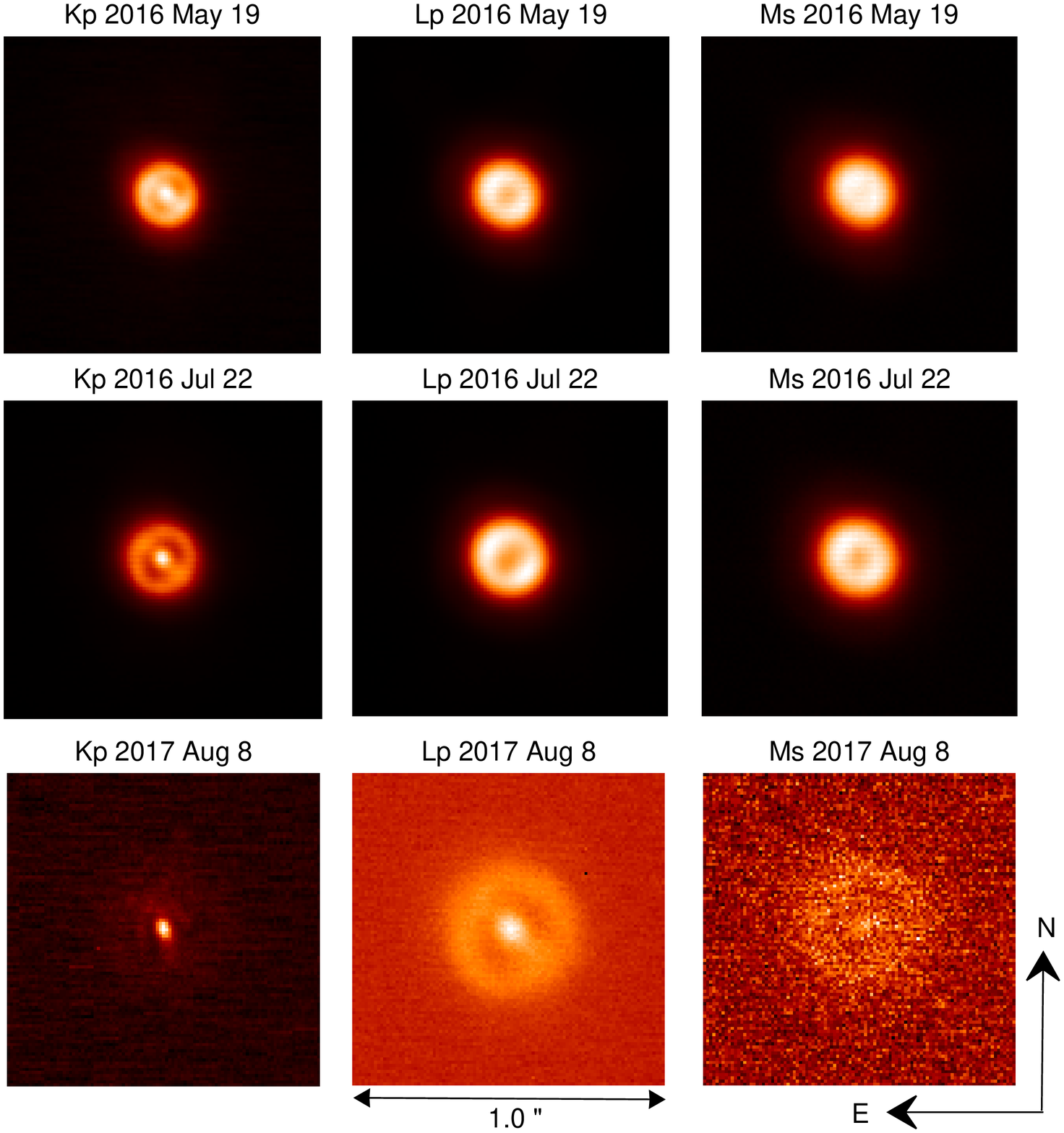}
    \caption{V5668 Sgr dust emission images in \textit{Kp} ($2.1$ \micron), \textit{Lp} ($3.8$ \micron) and \textit{Ms} ($4.7$ \micron) bands, in 2016 and 2017. The dust was not detected in 2019 data.}
    \label{fig:nirc2}
\end{figure*}

V5668 Sgr was observed with the the LGSAO assisted instrument on Keck II, the Near InfraRed Camera-2, NIRC2, in May of 2016 ($\Delta$t = 431 days), in July of 2016 ($\Delta$t = 495 days), August of 2017 ($\Delta$t = 876 days), and August of 2019 ($\Delta$t = 1613 days). NIRC2 has sensitivity in the thermal IR out to $5$ \micron\ and this capability was used to map the dust emission in the ejecta of V5668 Sgr as show in Figure \ref{fig:nirc2}. The warm dust emitted brightly with a black-body equivalent temperature of ~500 K in 2016, faded and cooled significantly by 2017 to less than 400 K, and was not detected in 2019 in the \textit{L-prime} ($3.8$ \micron) and \textit{Ms} ($4.7$ \micron) filter band-passes. The black-body equivalent temperatures were based on \textit{KLM} aperture photometry of a small section of the nova shell only, with the central source excluded from the aperture. Previous estimates of dust temperature of V5668 Sgr were made by \cite{2018ApJ...858...78G} for several dates until $\Delta$t $\sim$400 days. They found values ranging from 700 to 1100 K, with a temperature of 971$\pm$16 K at day 399. Their results may be overestimated because of the central source contribution, especially in the \textit{K} band, as we can see in NIRC2 images. The high spatial resolution AO data enables our measurements to distinguish the flux in the ejecta from that of the central source. The nebular expansion and morphology of the dust matches closely with that of the gas as resolved in the HST and OSIRIS data.

\section{Expansion parallax}

\begin{figure}
    \centering
    \includegraphics{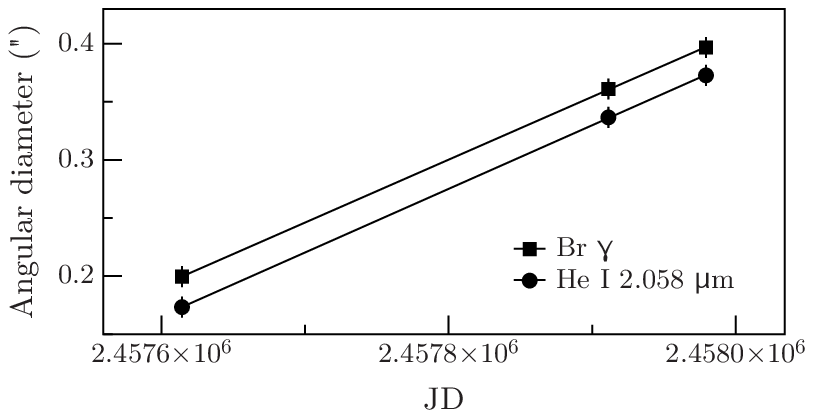}
    \caption{V5668 Sgr angular diameter expansion fit to Br$\gamma$ and \ion{He}{i} 2.058~\micron\ OSIRIS images. The obtained expansion rate of 0\farcs20 $\pm$ 0\farcs04 per year lead to a distance of 1200 $\pm$ 400 pc. }
    \label{fig:expansion_rate}
\end{figure}

Using OSIRIS data from the three epochs, we were able to estimate the expansion rate of the torus component, using Br$\gamma$ and \ion{He}{i} 2.058~\micron\ lines. The torus shape is elliptical due to projection effects, therefore we used average values of the major and minor axes measured at the line's peak intensity. For both Br$\gamma$ and \ion{He}{i} features, we measured an expansion rate for the angular diameter of $0.20$~arcsec\,year$^{-1}$ (Figure \ref{fig:expansion_rate}). The same analysis could not be applied to the polar caps structures because the blueshift and redshift components overlap near the center, making it difficult to determine the angular distance between the caps.

\begin{table}
    \centering
    \caption{Radial velocities of the polar component (km\,s$^{-1}$)}
    \begin{tabular}{c|c|c|c|c}
    Date & unk $2.1$ \micron  & [\ion{Si}{vi}] & Br$\gamma$ & \ion{He}{i} $2.058$ \micron \\
    \hline
      2016-08-13   & 567 & 536 & 548 & 521\\
      2017-06-06   & 518 & - & 557 & 545\\
      2017-08-13   & 535 & 552 & 516 & 545
    \end{tabular}
    \label{tab:radvel_polar}
\end{table}

\begin{table}
    \centering
    \caption{Radial velocities of the equatorial component (km\,s$^{-1}$)}
    \begin{tabular}{c|c|c}
    Date &  Br$\gamma$ & \ion{He}{i} $2.058$ \micron \\
    \hline
      2016-08-13   & 262 & 227\\
      2017-06-06   & 247 & 236\\
      2017-08-13   & 231 & 228
    \end{tabular}
    \label{tab:radvel_eq}
\end{table}

The lack of detected eclipses and orbital velocity for V5668 Sgr may indicate that it is a low-inclination system. This scenario favours the interpretation of a spherical expansion, in which the expansion velocity is isotropic. The assumption of a spherical expansion may introduce errors in the distance determination \citep{2000PASP..112..614W}, especially when there are no evidences of the shell geometry, but the results of the photoionization models described in section \ref{sec:rainy} corroborate an approximately spherical shell rather than a prolate one. The dust distribution observed in NIRC2 images, with bright polar emission in 2016, also indicates that the matter is not expanding faster in the poles, otherwise we would expect to have a lower grain density in these regions. 

Assuming a spherical expansion, the radial velocities are given by the projections of the expansion velocity in different angles. Therefore, we measured the radial velocities of the polar caps for four emission lines in all epochs (Table \ref{tab:radvel_polar}), obtaining a mean value of radial velocity v$_{\rm pol}=540$ $\pm$ 16 km\,s$^{-1}$. The values in the table correspond to the highest value of Doppler displacement measured in the blueshift and redshift peaks of the lines, and the error corresponds to the standard deviation of the measurements. We considered a uniform expansion velocity for all of our observations, with temporal variations attributed to instrumental uncertainties rather than to acceleration and deceleration of the gas. We repeated the analysis for the equatorial component (Table \ref{tab:radvel_eq}), with the highest velocity from the single peaked component of the lines, for which we estimated v$_{\rm eq}=238$ $\pm$ 14 km\,s$^{-1}$. These values lead to an inclination angle of $24^{\circ}$ and an expansion velocity of $590$ $\pm$ 18 km\,s$^{-1}$. This velocity is slightly higher than the expansion velocity of $530$ km\,s$^{-1}$ assumed by \cite{2016MNRAS.455L.109B} by measuring the HWHM of Br$\gamma$ line in a 2015 1D spectrum.

Correcting the projection effect for the expansion rate measured for the angular diameter, we obtained a radial expansion rate of $0.10$ arcsec\,year$^{-1}$. By combining the expansion velocity and the expansion rate, we derived a distance of $1.2$ $\pm$ 0.4 kpc. \cite{2018ApJ...858...78G} found the same distance value of $1.2$ kpc from MMRD relation, while \cite{2016MNRAS.455L.109B} estimated a distance of $1.54$ kpc assuming a blackbody angular diameter and the previously cited expansion velocity. \cite{2021ApJ...910..134G} found a larger value for the distance of $2.8 \pm 0.5$ kpc using their extinction estimates allied to a 3D Galactic reddening map \citep{2019MNRAS.483.4277C}, with the uncertainty in the 3D dust distribution being the likely source of the discrepancy. By the date of this writing, there were no Gaia data available for this object.

\section{Hydrogen density map} \label{sec:3Dshell}

\begin{figure}
    \centering
    \includegraphics[width=4cm]{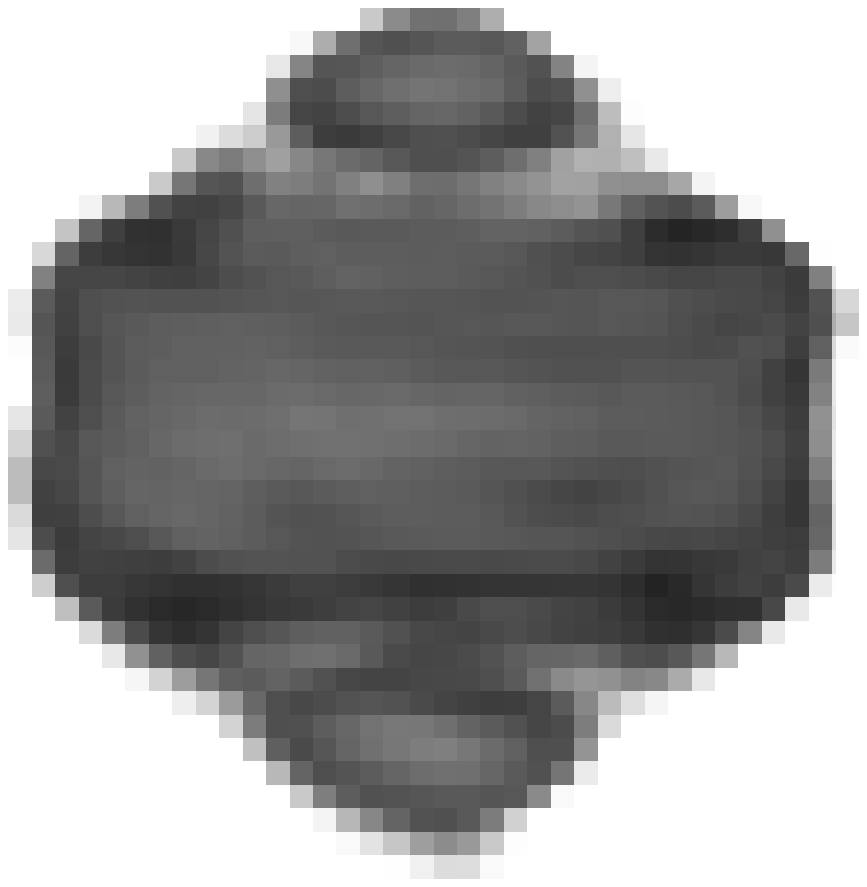}
    \includegraphics[width=4cm]{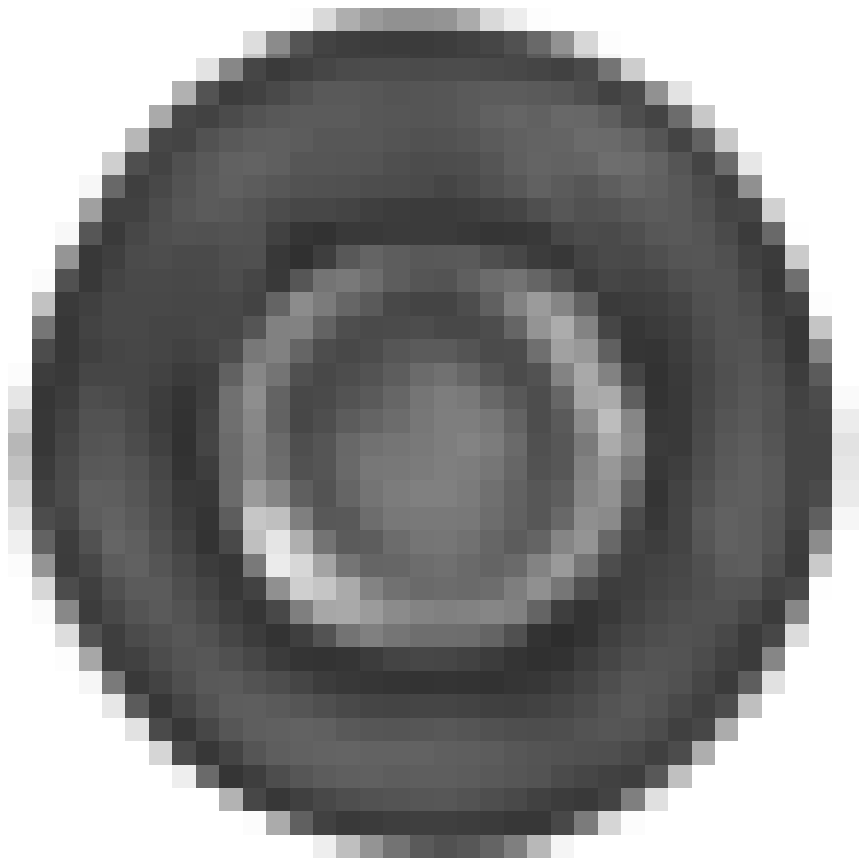}
    \caption{3D hydrogen density map, with denser regions represented in darker colour. The left image is a visualisation of the projection in the polar plane, and the right image is the visualisation of the projection in the equatorial plane.}
    \label{fig:hden}
\end{figure}

Once we assumed a spherical expansion and estimated the distance, we were able to convert the spectral axis into a third position axis, building a 3D hydrogen emission map from the Br$\gamma$ data cube. We considered ionization equilibrium recombination \citep{osterbrock2006astrophysics} in order to convert the line emission map into a density distribution, assuming optically thin lines. The result for August 13 2017, the latest date of OSIRIS data, is displayed in Figure \ref{fig:hden}, with two views of the 3D shell. The resulting density map is converted into a spherical grid, with 70 radial steps, 9 polar angle steps and a maximum of 8 azimuthal angle steps. 

It is important to stress that the observed geometry of the shell from imaging is usually assumed to correspond directly to the gas distribution, but distinct processes generate distinct emissivity functions of density. For the recombination process of hydrogen, for instance, the emissivity scales approximately with the squared density, which explains the differences between our density grid and the hydrogen images. Another caveat is that the neutral gas is neglected, which can affect the mass estimates especially in the presence of clumps.

\section{Abundances}\label{sec:abund}

The NIR spectra of V5668 Sgr exhibit numerous bright \ion{He}{i} emission lines, which are especially important for constraining the photoionization models in the presence of \ion{H}{i} recombination. The theoretical recombination flux ratio between \ion{He}{i} 2.058 \micron\ and Br$\gamma$ \citep{1987MNRAS.224..801H} was used to infer the He abundance because other \ion{He}{i} and \ion{He}{ii} lines seem blended. This procedure was applied to all Keck-OSIRIS line fluxes and also to the NIR spectra fluxes published by \cite{2018ApJ...858...78G}. We noticed a decrease in this flux ratio with time, suggesting the presence of significant neutral He in the later phases. Therefore we adopted the highest (and earlier in time) flux ratio, that lead us to n$_{\rm He}/$n$_{\rm H}=0.45$. This is a high He abundance for a nova (\citealt{1998PASP..110....3G}; while even more extreme values are found in the literature, \citealt{2001MNRAS.320..103S}), suggesting the presence of a He-rich companion. 

As mentioned above, V5668 Sgr presented measurable $^{7}$Be in the early spectra. $^{7}$Be is produced during the thermonuclear runaway, in the reaction $^{3}$He($\alpha$,$\gamma$)$^{7}$Be, and decays through electron capture into $^{7}$Li after a period of 53.2 days \citep{2003NuPhA.729....3A}. \cite{Molaro} derived N($^{7}$Be)/N(Ca) $\sim 53-69$, which leads to N(Li) $=4.7-4.9$ N(Li)$_{\sun}$, while \cite{2016ApJ...818..191T} found a lower abundance ratio of N($^{7}$\ion{Be}{ii})/N(\ion{Ca}{ii}) $\sim 8.1$, or X($^{7}$Be)/X(Ca) $= 1.4$. However, recent discussions about the calculation of $^{7}$Be abundance through absorption lines indicate that these values may be overestimated \citep{2020AstL...46...92C}. We adopted the ratio of N($^{7}$\ion{Be}{ii})/N(\ion{Ca}{ii}) $\sim 8.1$ as an upper limit, although our spectra do not show any Li emission lines in order to directly evaluate this abundance. 

The carbon abundance was set assuming that most of the carbon dust estimated by \cite{2018ApJ...858...78G} was completely depleted by the time of our models, which could be slightly overestimated since we still observe dust in 2017 NIRC2 images. On the other hand, considering that the gas carbon abundance is expected to be higher than the grains' abundance, our estimate is probably a lower limit. For N and O abundances, average C, N and O for novae were used to calculate the N/C and O/C ratios. These values were scaled with the estimated carbon abundance, leading to log(n$_{\rm C}/$n$_{\rm H})=-2.6$, log(n$_{\rm N}/$n$_{\rm H})=-3.7$ and log(n$_{\rm O}/$n$_{\rm H})=-3.1$. The resulting O/C ratio is lower than 1, which is compatible with the formation of amorphous carbon dust \citep{2018ApJ...858...78G}. All the other metal abundances including Ne were set as solar values \citep{Asplund}.

\section{\textsc{rainy3d} photoionization models} \label{sec:rainy}

The photoionization models were performed with the code \textsc{rainy3d} \citep{Rainy}, which runs \textsc{cloudy} \citep{Cloudy} as a subroutine. \textsc{rainy3d} is capable of describing the local and integrated spectrum for an arbitrary mass distribution with the approximation of radial radiative transfer. The parallelized calculations presented here were performed at the Santos Dumont 5 petaflop supercomputer facility. We used HST data from July 2017 and OSIRIS data from August 2017 for the modelling, in order to combine the optical and NIR information of the same epoch, considering that the central source would not have significantly cooled down in one month. The main input parameters are described in Table \ref{tab:input_phot}. We varied the physical properties of the central source, namely the temperature and luminosity, over the extended value ranges found in classical novae. Unfortunately, there are no contemporary data in X-rays or EUV for constraining the ionizing source properties at the time. We have adopted the Rauch NLTE hot high-gravity stellar atmosphere grid \citep{2003A&A...403..709R} for the central source SED. 

The shell mass was varied around the recombination mass estimated from the integrated hydrogen density grid. Although that was varied, we fixed the density gradient as the one described in section \ref{sec:3Dshell}. We sampled the shell mass in 7 steps within the interval of $3.0\times 10^{-5}$ to $3.0 \times 10^{-4}$ M$_{\sun}$, the central source temperature in 7 steps within the range of 80,000 to 210,000 K, and the central source luminosity in 5 steps from $1\times 10^{35}$ to $1\times 10^{36}$ erg\,s$^{-1}$. 

\begin{table}
    \centering
    \caption{Input values for the shell and the central source (CS) used in the 3D photoionization models.}
    \begin{tabular}{l|c}
      Distance (pc)    & $1200$  \\
      \textit{E(B-V)}     & $0.2$ \\
      Temperature$_{\rm CS}$ (K)     & $(80 - 210)\times 10^{3}$\\
      Luminosity$_{\rm CS}$ (erg\,s$^{-1}$)    & $1\times 10^{35} - 1\times 10^{36}$ \\
      Mass$_{\rm shell}$ (M$_{\sun}$)    & $3.0\times 10^{-5} - 3.0 \times 10^{-4}$ \\
      r$_{\rm in}$ (cm) & $5.2\times 10^{14}$\\
      r$_{\rm out}$ (cm) & $5.2\times 10^{15}$\\
    \end{tabular}
    \label{tab:input_phot}
\end{table}

A list of observed emission lines fluxes are compared with the integrated fluxes obtained from the models. We also included emission lines typically observed in the NIR region of nova spectra but not observed in our Keck-OSIRIS data in order to constrain our models. The values are displayed in Table \ref{tab:line_list}. The NIR lines fluxes presented as upper limits are either lines that are not observed or lines possibly blended with unidentified transitions. Based on the HST filter widths and the expansion velocity, we estimated that the F657N total flux would be an upper limit for the combination of H$\alpha$, [\ion{N}{ii}] $\lambda\lambda$ 6548.05 \AA\,  6583.45 \AA\ and \ion{He}{ii} 6559.91 \AA\ fluxes. On the other hand, the integrated flux from F502N HST image should be dominated only by the [\ion{O}{iii}] 5006.84 \AA\ flux. 

\begin{table}
    \centering
    \caption{List of integrated emission lines fluxes (in erg\,s$^{-1}$\,cm$^{-2}$), with wavelength values based on \textsc{cloudy} atomic data.}
    \begin{tabular}{l|c}
{[\ion{N}{ii}]} 6548.05 \AA\ + \ion{He}{ii} 6559.91 \AA\   & \multirow{2}{*}{$\leq4.2\times 10^{-11}$} \\
+ H$\alpha$ 6562.81 \AA\ + [\ion{N}{ii}] 6583.45 \AA & \\
{[\ion{O}{iii}]}  5006.84 \AA &  $1.05\times 10^{-10}$ \\
{[\ion{Si}{vi}]} 1.96247 \micron &  $1.0\times 10^{-13}$ \\
\ion{He}{ii} 2.03725 \micron &  $\leq1\times 10^{-14}$ \\
\ion{He}{i} 2.05813 \micron & $1.37\times 10^{-13}$ \\
\ion{He}{i} 2.11303 \micron & $\leq2.2\times 10^{-13}$ \\
Br$\gamma$ 2.16551 \micron & $5.3\times 10^{-13}$ \\
\ion{He}{ii} 2.18843 \micron & $\leq1.2\times 10^{-13}$ \\
{[\ion{Ca}{viii}]} 2.32117 \micron & $\leq1\times 10^{-14}$ \\
\ion{He}{ii}    2.34631 \micron & $\leq1\times 10^{-14}$ \\
    \end{tabular}
    \label{tab:line_list}
\end{table}

In order to analyse how an anisotropic ionizing field would influence the observed morphologies, we also performed models with an accretion disc. The geometrically thin, optically thick, multi-temperature standard disc \citep{frank2002accretion} was aligned to the equatorial region in a this model grid. A very simplistic model of the disc is considered. For instance, it does not consider scattering of photons from the central source, limb darkening effects and the disc flare. In our estimates for the disc luminosity and SED, we have assumed a white dwarf mass of 1.1 M$_{\sun}$ \citep{2018ApJ...858...78G}, which has a linear effect in the disc luminosity and temperature, and we varied the mass transfer rate in a wide interval from $1\times10^{-9}$ to $5\times10^{-7}$ M$_{\sun}$\,year$^{-1}$. We also varied the spherical central source luminosity from $10^{34}$ to $10^{36}$ erg\,s$^{-1}$, but we fixed the other physical and chemical parameters of the shell to the best-fitting values from previous models.

\section{Results and discussion} \label{sec:discussion}

\begin{figure}
    \centering
    \includegraphics{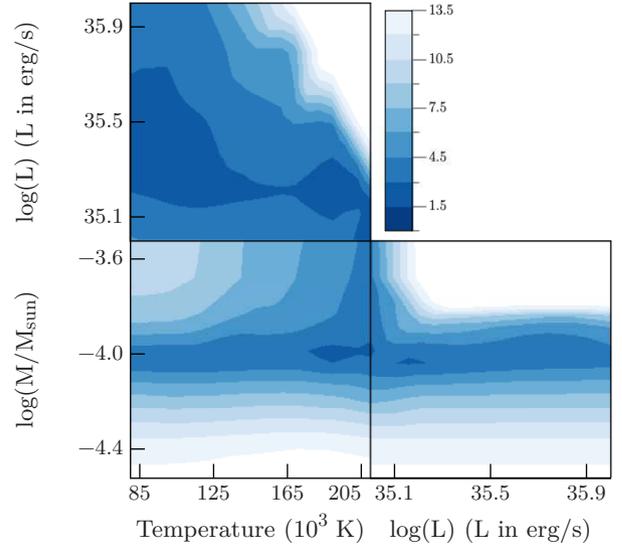}
    \caption{$\chi^2$ grid of the photoionization models as function of total shell mass, central source temperature and its luminosity.}
    \label{fig:chisqr}
\end{figure}

The emission line fluxes predicted by our model grid with isotropic ionization are shown in Figure \ref{fig:chisqr}, as contour maps of weighted and reduced $\chi^{2}$ as functions of the shell mass and central source luminosity and temperature. The minimum $\chi^{2}$ values, related to the dark-colored regions, point towards a best-fitting solution with T=188,000~K, L=1.6$\times10^{35}$ erg\,s$^{-1}$ and M=6.3$\times10^{-5}$ M$_{\sun}$. Assuming the $\chi^2$ valley widths around minimum values as upper limits for the uncertainties, we obtain uncertainties of $1.0\times10^{35}$ erg/s for the luminosity and $30,000$ K for the temperature of the central source. We note that the shell mass has more influence in the results than the central source parameters. The estimated shell mass is compatible with other values previously obtained by other authors using different data and methods, such as M=2.4$\times10^{-5}$ M$_{\sun}$ \citep{2018ApJ...858...78G} and M=$2.7-5.4\times10^{-5}$ M$_{\sun}$ \citep{2016MNRAS.455L.109B}. The derived central source luminosity is compatible with the luminosity of a hot pseudo-photosphere on a white dwarf with $\sim$1.1 M$_{\sun}$, as was also found by \cite{2018ApJ...858...78G}.

In our best-fitting model, all line fluxes and line flux upper limits are compatible within a 30 per cent uncertainty in absolute fluxes, with the exception of Br$\gamma$ and [\ion{Si}{vi}]. The modelled Br$\gamma$ and [\ion{Si}{vi}] fluxes are a factor of 5 and 2.5 below the observed fluxes respectively. The direct increase in the ionization level would brighten the non-detected [\ion{Ca}{viii}] line, and as we already varied the shell mass, one could suggest that the chosen abundances may be wrong. In fact, when we slightly change the Si abundance from the solar value of log(n$_{\rm Si}$/n$_{\rm H}$)~=-4.49 to log(n$_{\rm Si}$/n$_{\rm H}$)~=-4.2, the [\ion{Si}{vi}] model flux becomes compatible with the observed one. The increase of Br$\gamma$ flux can also be achieved by lowering the He abundance relative to H in the shell, which increases the H mass fraction. When we lowered the He abundance to log(n$_{\rm He}$/n$_{\rm H}$)~=-0.7, maintaining the modelled \ion{He}{i} and \ion{He}{ii} line fluxes compatible with the observations, we were able to obtain a higher Br$\gamma$ flux of $1.5\times10^{-13}$, which is a better fit but still lower than the observed flux of $5.3\times10^{-13}$~erg\,s$^{-1}$\,cm$^{-2}$. The exact carbon and oxygen abundances are also unknown and can affect the model's results. As described in section \ref{sec:abund}, we used a  carbon abundance corresponding to the estimate of the maximum carbon dust mass produced in the ejecta \citep{2018ApJ...858...78G}. The increase of this abundance would favour the cooling processes of the gas, and would not contribute to the increase of H or Si fluxes in the models.

\begin{figure}
    \centering
    \includegraphics[width=4cm]{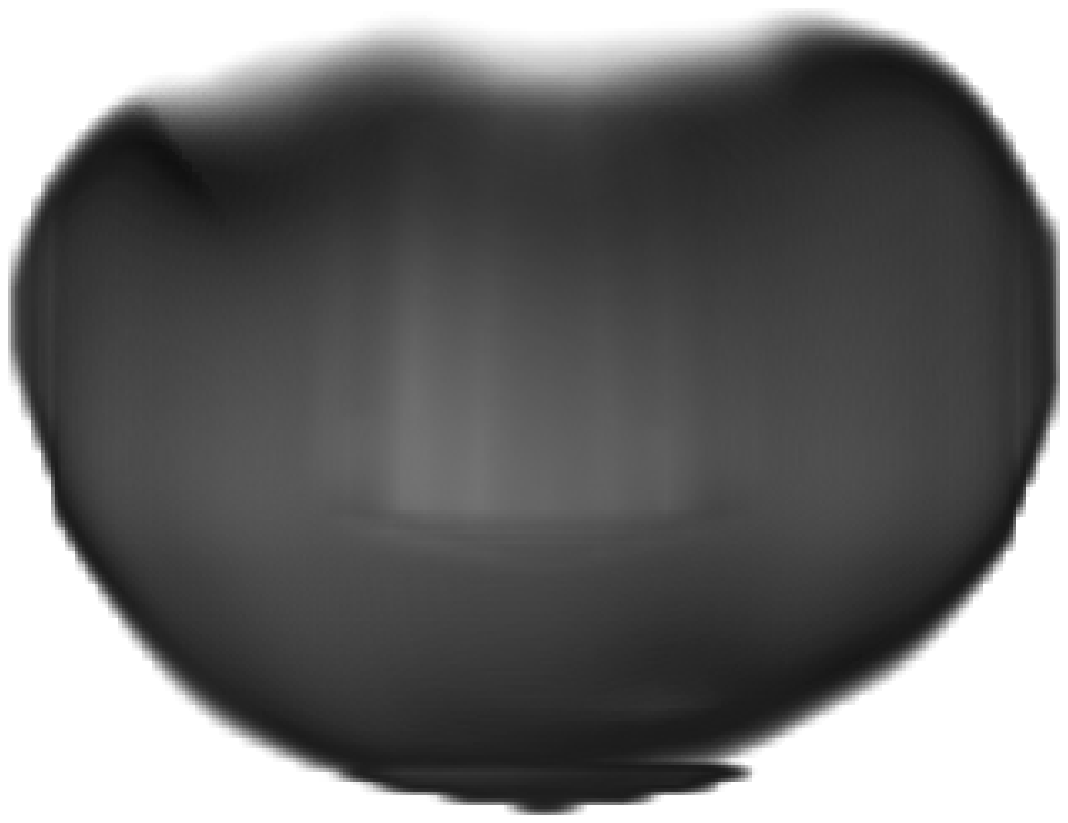}
    \includegraphics[width=4cm]{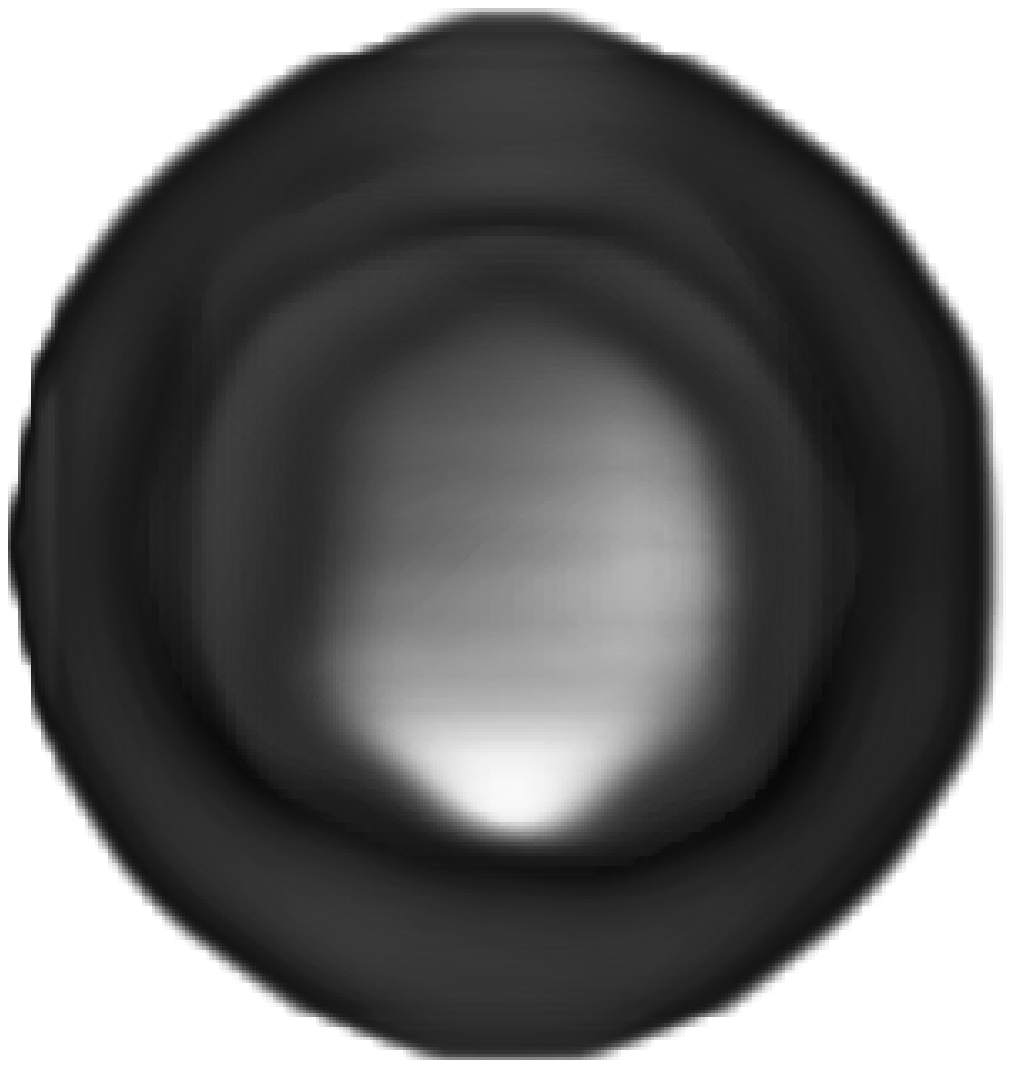}
    \caption{3D Br$\gamma$ emissivity map. The left image is a  visualisation of the projection in the polar plane, and the right image is the visualisation of the projection in the equatorial plane.}
    \label{fig:hmod}
\end{figure}

\begin{figure}
    \centering
    \includegraphics[width=4cm]{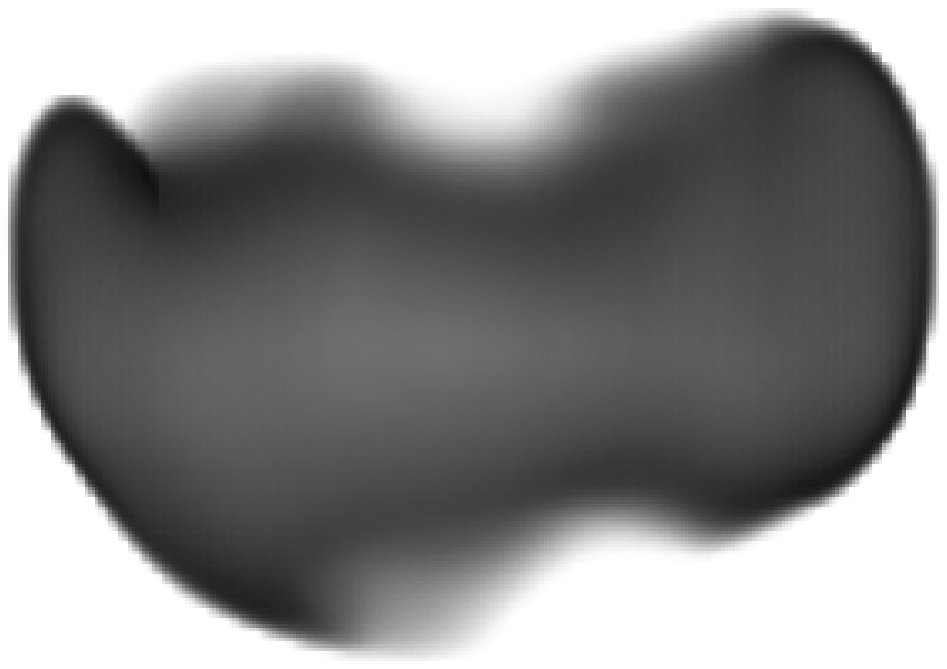}
    \includegraphics[width=4cm]{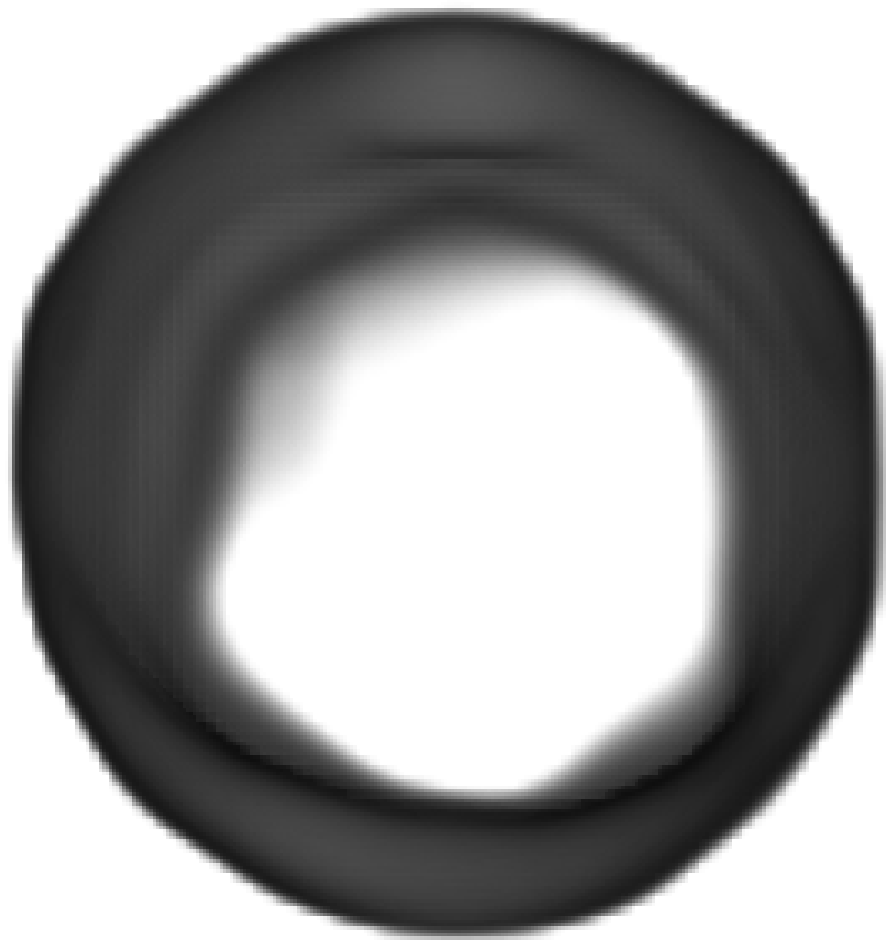}
    \caption{3D [\ion{Si}{vi}] emissivity map. The left image is a  visualisation of the projection in the polar plane, and the right image is the visualisation of the projection in the equatorial plane.}
    \label{fig:simod}
\end{figure}

Although we obtained a reasonable fitting of total emitted fluxes in these models, we could not achieve the difference in the morphology displayed by [\ion{Si}{vi}] and Br$\gamma$ lines. For both lines, the 3D models show the equatorial region more prominent than the polar caps (see Figures \ref{fig:hmod} and \ref{fig:simod}). For Br$\gamma$ emissivity map, we note that the shell is slightly asymmetric, with the lower half brighter than the upper half. For [\ion{Si}{vi}] however, the modelled morphology is not compatible with the observation. These results indicate that the density gradient is not enough to explain the distinct structures in the nova shell. 

\begin{figure}
    \centering
    \includegraphics[width=4cm]{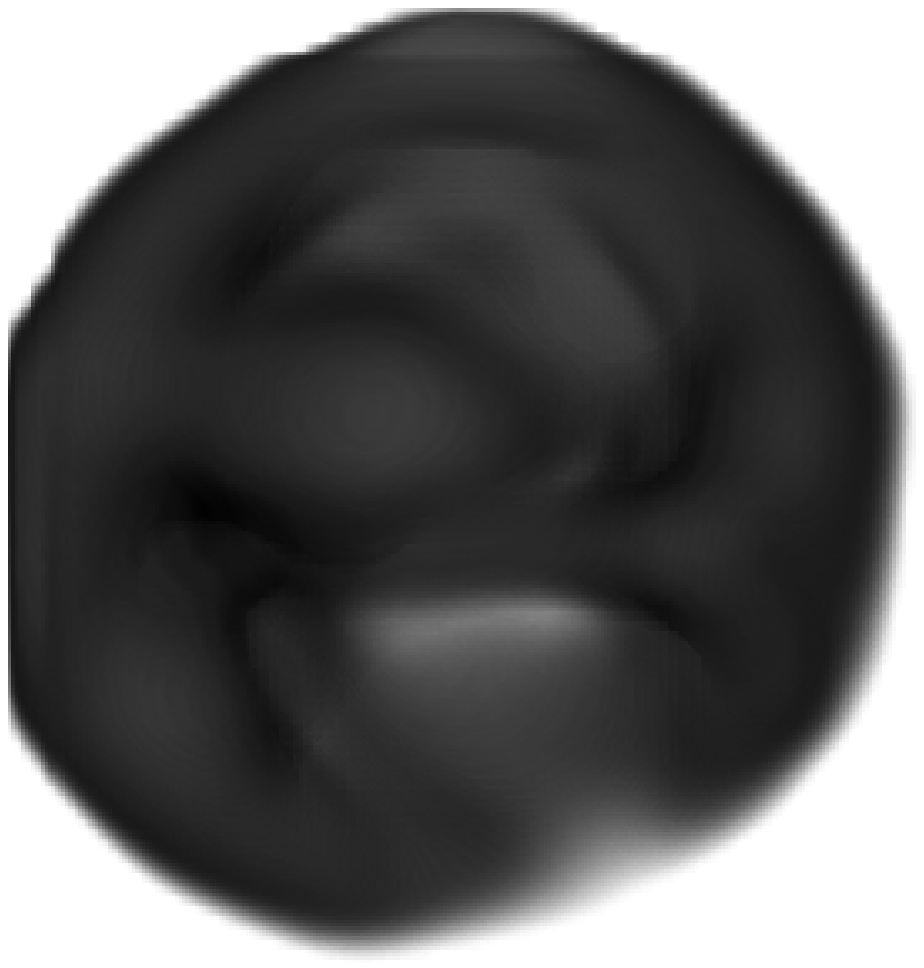}
    \includegraphics[width=4cm]{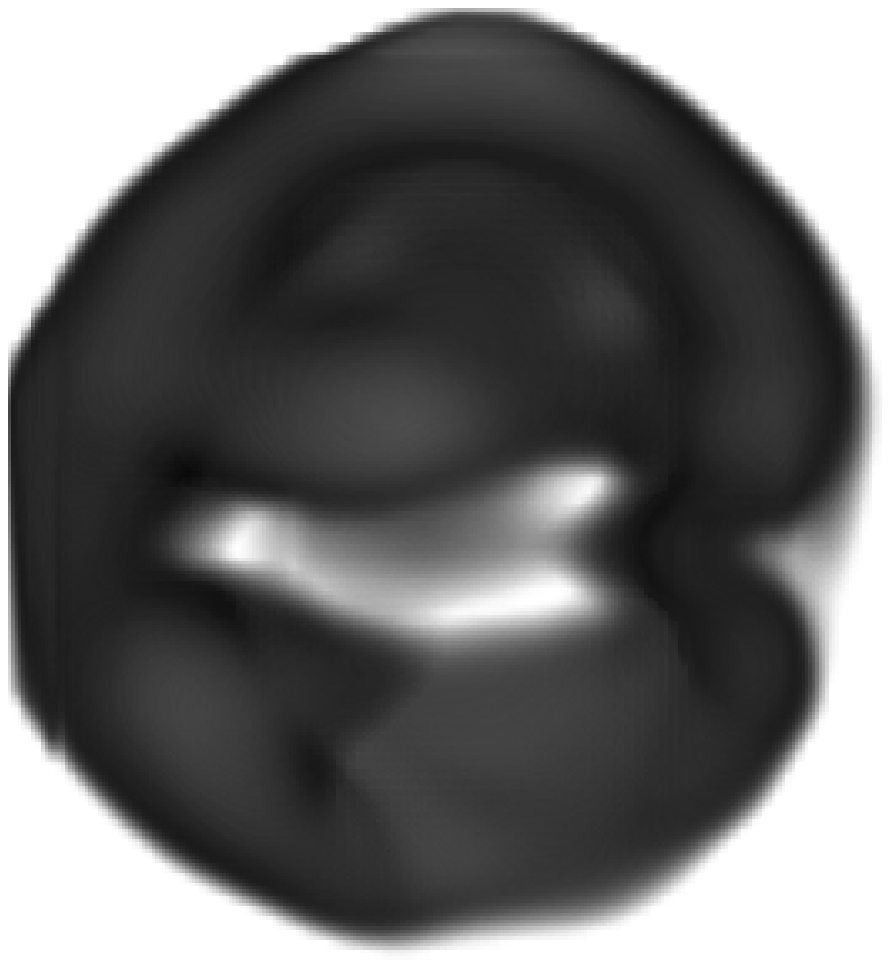}
    \caption{Visualisation of Br$\gamma$ (left) and [\ion{Si}{vi}] (right) emissivity maps for a model with accretion disc.}
    \label{fig:hdisk}
\end{figure}

In our models with anisotropic ionization generated by an accretion disc, we analysed the different morphologies produced by all luminosity weighted combinations for the ionizing source. We found that the bipolar aspect of [\ion{Si}{vi}] only starts to appear in the models where the accretion disc is more luminous than the central component, and it is best observed for the models with \.{M}=$5\times 10^{-7}$ M$_{\sun}$ year$^{-1}$ (Figure \ref{fig:hdisk}). However, an accretion disc with such a high mass transfer rate is hot and possibly thick at outer borders, resembling the discs  obscuring the direct view of the white dwarf in some persistent and post-nova supersoft sources \citep{2012ApJ...745...43N,2013A&A...559A..50N,2021arXiv210803241S}. Probably, its precise SED and vertical structure, and thus its effect on the shell ionization can not be predicted with our simplified standard disc assumptions \citep{frank2002accretion}. We would expect this disc to produce an enhanced anisotropy in the ionizing field due to its shape, significantly lowering the ionization parameter in the shell equatorial region. This effect was already observed in nova V723 Cas \citep{2009AJ....138.1090L}, in which the [\ion{Al}{ix}] line, exclusively polar, strengthens later in the shell development, perhaps as the disc reforms and hard UV is directed to the poles. Studies of novae, dwarf novae and supersoft sources show that accretion discs either survive the nova event or reform quickly, in timescales as short as 30 days \citep{Starrfield2004,1997MNRAS.286..745R,2002Sci...298..393H}.
 
The structures of the ionized shell observed in the lower luminosity models were not only incompatible with the highly ionized polar caps, but were also similar to an oblate spheroid since the ionization parameter rapidly decays with the distance to the gas.

A reestablished luminous accretion disc has been considered as part of the ionizing source in other novae, such as V723 Cas \citep{TakedaV723Cas} and HR Del \citep{2009AJ....138.1541M}. In the case of V723 Cas the disc was found to be crucial in describing the observed ionization structure of the shell.

Alternatively to the presence of a disc, one could also try to explain the different geometries observed for the different transitions in Keck data with a non-uniform distribution of chemical elements within the shell. However, due to the mixing and convection processes during the eruption, this scenario seems unlikely to occur.

The three-dimensional analysis of nova remnants highlights the difficulties in interpreting the observed structures in the shell. The influences of mass distribution, abundance gradient and ionizing field in the geometry of the ionized gas are not fully understood, nor are the processes responsible for them. The combination of the eruption process, shocks, winds and interaction with pre-existing circumstellar gas could explain a large variety of shell structures that could be studied using 3D hydrodynamic simulations. 

In the case of V5668 Sgr as a nova that had detected gamma-ray emission in the early phase of eruption, the analysis of the shock is especially important for the understanding of the gas dynamics and the shell structure. The amplification of magnetic fields in internal \citep{2014Natur.514..339C} or external shocks \citep{2010Sci...329..817A} is the currently accepted mechanism for explaining non-thermal radio emission from novae. Its observed fast variability and flaring phenomena \citep{2021MNRAS.501.1394N} suggests rapidly changing of shear velocities and/or density contrasts in the shocking flow, leading to a complex anisotropic emission scenario. A relation of such variability in synchrotron emission and peculiar clump velocities inside the shell has yet to be found. Precise synoptic astrometry of condensations combined with high resolution 2D spectroscopy may be able to address the formation of shocks associated with runaway clumps. On the other hand, secondary eruptions and the pre-existing circumbinary gas distribution are also correlated with the non-thermal temporal behaviour.

\section{Conclusions}  \label{sec:conclusions}

We present HST optical images and Keck-OSIRIS and Keck-NIRC2 NIR data from the evolved remnant of nova V5668 Sgr. The IFS data permitted the estimate of a parallax distance of 1200 $\pm$ 400 pc, assuming an isotropic expansion velocity that lead to v=590 km\,s$^{-1}$ and a system inclination angle of 24$\degr$. 

The observed gas structures in all data are different depending on the ionization of the emission line. For the highly ionized transitions, the shell presents enhanced polar caps, while for the lower ionization lines the shell also presents strong equatorial emission. Our photoionization models suggest that this anisotropy of the ionizing field may be due to the presence of a luminous reestablished accretion disc. The dust distribution follows the gas distribution between 2016 and 2017, with black-body temperatures close to $\sim500$~K. We could not detect dust emission in 2019.

Our best-fitting models for August of 2017 indicate that the central source has a temperature of $188,000$~K and a luminosity of $1.6\times10^{35}$ erg\,s$^{-1}$, and that the shell has a total ejected mass of $6.3\times10^{-5}$ M$_{\sun}$. Even though we were able to reproduce the observed integrated line fluxes, we could not obtain the observed shell structures, possibly due to the limitations of our accretion disc modelling.

\section*{Acknowledgements}

We thank S\~ao Paulo Research Foundation (FAPESP) for the support under grant 2019/08341-8 and CNPq funding under grant \#305657. J.D.L. was supported in part by grant HST-GO-14787.010-A. We also acknowledge with thanks the variable star observations from the AAVSO International Database contributed by observers worldwide and used in this research. The authors acknowledge the National Laboratory for Scientific Computing (LNCC/MCTI, Brazil) for providing HPC resources of the SDumont supercomputer, which have contributed to the research results reported within this paper. URL: http://sdumont.lncc.br

This research is based, in part, on observations made with the NASA/ESA {\it Hubble Space Telescope} obtained from the Space Telescope Science Institute, which is operated by the Association of Universities for Research in Astronomy, Inc., under NASA contract NAS 5-26555. These observations are associated with program 14787.

This publication makes use of data products from the Two Micron All Sky Survey, which is a joint project of the University of Massachusetts and the Infrared Processing and Analysis Center/California Institute of Technology, funded by the National Aeronautics and Space Administration and the National Science Foundation.

Some of the data presented herein were obtained at the W.~M.~Keck Observatory, which is operated as a scientific partnership among the California Institute of Technology, the University of California and the National Aeronautics and Space Administration. The Observatory was made possible by the generous financial support of the W.~M.~Keck Foundation.

The authors wish to recognize and acknowledge the very significant cultural role and reverence that the summit of Maunakea has always had within the indigenous Hawaiian community.  We are most fortunate to have the opportunity to conduct observations from this mountain.

The work of K.V.S. was supported by the National Science Foundation under Grant~No.~AST-1751874 and NASA grants NASA/NuSTAR 80NSSC21K0277, NASA/Fermi 80NSSC20K1535, and NASA/Swift 80NSSC21K0173 and from a Cottrell Scholarship from the Research Corporation. We acknowledge support from the Packard Foundation. 

We would like to thank the reviewer, whose comments and suggestions were important to improve our manuscript. We would also like to acknowledge V. A. R. M. Ribeiro for his useful contributions to the discussion of the results.

%%%%%%%%%%%%%%%%%%%%%%%%%%%%%%%%%%%%%%%%%%%%%%%%%%
\section*{Data Availability}

The NIR data underlying this article are available in Keck Observatory Archive (KOA) at \url{https://koa.ipac.caltech.edu/}, and can be accessed with program IDs K386, K253, K281OL, K214N2L. The HST optical data are available in MAST at \url{https://mast.stsci.edu/search/hst/ui/#/}, and can be accessed with the proposal ID 14787.



\bibliographystyle{mnras}
\bibliography{bibliography} % if your bibtex file is called example.bib

%\begin{thebibliography}{99}
%\end{thebibliography}

%%%%%%%%%%%%%%%%%%%%%%%%%%%%%%%%%%%%%%%%%%%%%%%%%%

%%%%%%%%%%%%%%%%% APPENDICES %%%%%%%%%%%%%%%%%%%%%

%\appendix

%%%%%%%%%%%%%%%%%%%%%%%%%%%%%%%%%%%%%%%%%%%%%%%%%%

% Don't change these lines
\bsp % typesetting comment
\label{lastpage}
\end{document}